\documentclass[reprint, superscriptaddress, secnumarabic, amssymb, nobibnotes, aps, prl]{revtex4-1}

\usepackage{graphicx}
\usepackage{epstopdf}
\usepackage[T1]{fontenc}
\usepackage[latin9]{inputenc}
\usepackage{amsbsy}
\usepackage{gensymb}
\usepackage{textcomp}
\usepackage[T1]{fontenc}
\usepackage[latin9]{inputenc}
\usepackage{amsmath}
\usepackage{amssymb}
\usepackage{bbm}
\usepackage{soul}
\usepackage{physics}
\usepackage{braket}
\usepackage{xcolor}
\usepackage{float}
\allowdisplaybreaks
\usepackage{graphicx}
\usepackage{array}
\usepackage[colorlinks=true]{hyperref}  
\hypersetup{
    unicode=false,          
    pdftoolbar=true,        
    pdfmenubar=true,        
    pdffitwindow=false,     
    pdfstartview={FitH},    
    pdftitle={Superconductivity in Equimolar High Entropy Alloy \texorpdfstring{Sc$_{20}$V$_{20}$Ti$_{20}$Hf$_{20}$Nb$_{20}$}{}},    
    pdfauthor={, , ,},     
    pdfsubject={},   
    pdfcreator={},   
    pdfproducer={},  
    pdfkeywords={} {} {}, 
    pdfnewwindow=true,      
    colorlinks=true,       
    linkcolor=blue, 
    citecolor=blue,        
    filecolor=magenta,      
    urlcolor=blue           
}
\usepackage[normalem]{ulem}

\newcommand{\equref}[1]{Eq.~(\ref{#1})}

\newcommand{\figref}[1]{Fig.~\ref{#1}}

\newcommand{\tableref}[1]{Table~\ref{#1}}

\begin{document}
\title{Superconductivity with high upper critical field in an equiatomic high entropy alloy Sc-V-Ti-Hf-Nb}

\author{S. Jangid}
\affiliation{Department of Physics, Indian Institute of Science Education and Research Bhopal, Bhopal, 462066, India}
\author{P. K. Meena}
\affiliation{Department of Physics, Indian Institute of Science Education and Research Bhopal, Bhopal, 462066, India}
\author{R. K. Kushwaha}
\affiliation{Department of Physics, Indian Institute of Science Education and Research Bhopal, Bhopal, 462066, India}
\author{S. Srivastava}
\affiliation{Department of Physics, Indian Institute of Science Education and Research Bhopal, Bhopal, 462066, India}
\author{P. Manna}
\affiliation{Department of Physics, Indian Institute of Science Education and Research Bhopal, Bhopal, 462066, India}
\author{P. Mishra}
\affiliation{Department of Physics, Indian Institute of Science Education and Research Bhopal, Bhopal, 462066, India}
\author{S. Sharma}
\affiliation{Department of Physics, Indian Institute of Science Education and Research Bhopal, Bhopal, 462066, India}
\author{R.~P.~Singh}
\email[]{rpsingh@iiserb.ac.in}
\affiliation{Department of Physics, Indian Institute of Science Education and Research Bhopal, Bhopal, 462066, India}

\begin{abstract}
High-entropy alloy (HEA) superconductors have attracted significant attention due to their exceptional low-temperature mechanical and superconducting properties. We report the synthesis and thorough characterization of an equiatomic HEA superconductor with the composition Sc$_{0.20}$V$_{0.20}$Ti$_{0.20}$Hf$_{0.20}$Nb$_{0.20}$, crystallizing in a body-centered cubic crystal structure (Im3$\bar{m}$). Our investigation, using magnetization, transport, and heat capacity measurements, reveals the presence of weakly coupled, fully gapped superconductivity with a transition temperature of 4.17(3) K and the upper critical field exceeding the Pauli paramagnetic limit. The metallic nature, combined with a high upper critical field, positions it as a promising candidate for applications in superconducting devices.
\end{abstract}
\maketitle

High-entropy alloys (HEAs) have gained attention as unconventional multicomponent alloys that contain five or more elements in a mixture \cite{yeh2004nanostructured}. High configurational entropy helps them stabilize disordered solid solutions in simple crystal structures akin to pure metals, including bcc, hcp, and fcc \cite{yeh2004nanostructured}. HEAs show remarkable properties such as high fracture toughness, hardness, strength, and exceptional corrosion resistance, making them valuable for a wide range of practical applications \cite{gludovatz2014fracture, zou2015ultrastrong, lee2007effect} such as structural and functional materials, magnetic
refrigeration, energy storage, radiation protection, bio-compatibility, and superconducting magnets in harsh environments \cite{funct, hstorage, radiation, biomat}. 

Recently, superconducting HEAs have emerged as a class of disordered alloy superconductors where the remarkable mechanical properties of HEAs are combined with the intriguing quantum phenomenon of superconductivity \cite{kovzelj2014discovery}. They have been reported to show remarkable superconducting properties, including retention of superconductivity under high pressure, elevated upper critical field, high critical current density, broadening in specific heat jump, and Debye temperature in the elemental range \cite{robustSCpressure, motla2023superconducting, jc, heathinfilm, kasem2021anomalous, marik2019superconductivity, motla2021probing}. In addition, its robust fracture strength at cryogenic temperatures and the potential to convert it into thin film form make them potential candidates for superconducting device application, particularly in extreme conditions \cite{jc, zhang2023superconductivity}. At the same time, its multicomponent nature offers a unique opportunity to tailor superconducting properties through precise composition and crystal structure adjustments. The highly disordered nature allows us to study the complex interplay of disorder and superconductivity. However, understanding the emergence of BCS superconductivity, even in the absence of conventional phonon modes required for conventional BCS superconductivity \cite{motla2022boron, motla2022superconducting, motla2023superconducting}, remains challenging, primarily due to limited studies on superconducting HEAs. Therefore, it is crucial to identify superconducting HEAs and characterize both normal and superconducting properties to advance our understanding of these disordered superconductors.

 Currently, substantial efforts have been made to find HEA superconductors with various crystal structures to explore their superconducting pairing mechanism and exotic properties \cite{status_cava, kitagawa2020cutting, von2016effect, strong2023superconductivity, stolze2018high, von2018isoelectronic, layered_SC, marik2018superconductivity, improved_layeredSC, ishizu2019new, marik2019superconductivity}. Here, we intend to explore the low-density equiatomic composition that yields the largest configurational entropy, thereby attaining the highest disorder. In this particular composition, each constituent element makes an equal contribution in determining the superconducting characteristics of the HEA. To achieve this, we have synthesized Sc$_{0.20}$V$_{0.20}$Ti$_{0.20}$Hf$_{0.20}$Nb$_{0.20}$ (ScVTiHfNb) primarily composed of 3d elements due to their lower density compared to the commonly utilized 4d/5d elements in superconducting HEAs. Among the chosen 3d elements, Sc, being the lightest, is known for its ability to enhance the mechanical properties of HEAs \cite{sun2023effects, ren2023sc}.

In this letter, we report the synthesis and characterization of an equiatomic bcc HEA ScVTiHfNb. Our results indicate that the alloy exhibits weakly coupled type II fully gapped superconductivity with a transition temperature of 4.17(3) K. Significantly, it demonstrates a remarkably high upper critical field that surpasses the Pauli limit, suggesting unconventional behavior. This unique feature, rarely observed in HEAs, makes it a promising candidate for applications in superconducting devices.

\begin{figure*}
\includegraphics[width=1.99\columnwidth, origin=b]{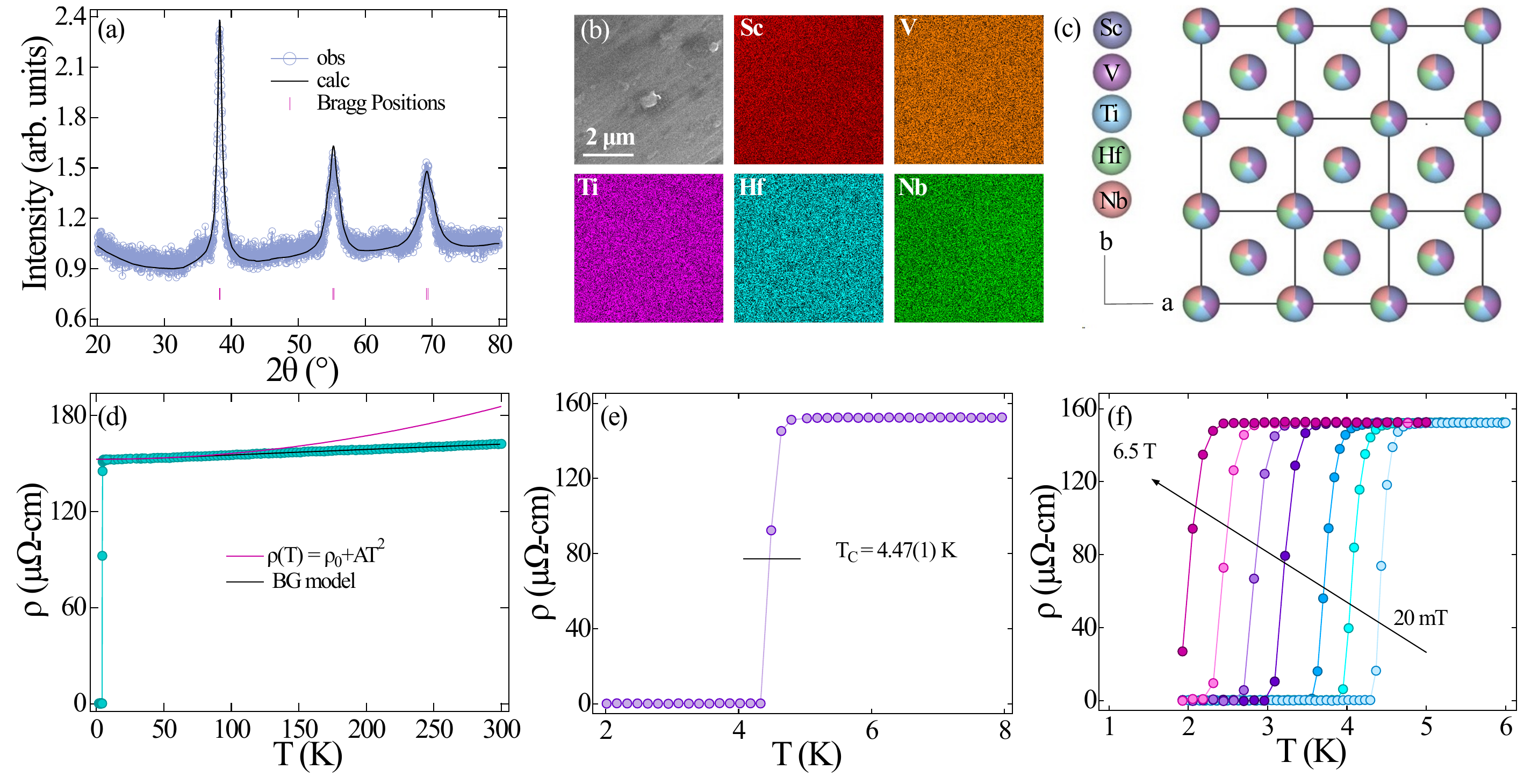}
\caption{\label{Fig1}(a) The powder X-ray diffraction of ScVTiHfNb shows crystallization in a bcc crystal structure (Im3$\bar{m}$). (b) Elemental mapping of Sc, V, Ti, Hf, and Nb elements. (c) Crystal structure of bcc ScVTiHfNb. (d) Temperature-dependent AC electrical resistivity up to room temperature in a zero field fitted with \equref{eqn1:BG} (solid black line) and power law (solid pink line). (e) Zero field electrical resistivity in the low-temperature range. (f) The resistivity vs. temperature curves under different applied magnetic fields.}
\end{figure*}
The polycrystalline sample of ScVTiHfNb was synthesized using a standard arc-melting technique. Stoichiometric amounts of high purity Sc (99.9$\%$), V (99.7$\%$), Ti (99.995$\%$), Hf (99.7$\%$), and Nb (99.8$\%$) pieces were melted at high currents in an argon atmosphere. The sample was flipped and remelted 5-6 times for increased phase homogeneity. Phase composition was confirmed by scanning electron microscope (SEM), while phase identification was performed by powder X-ray diffraction (XRD) using a PANalytical X$^{'}$Pert diffractometer equipped with CuK$_{\alpha}$ ($\lambda = 1.5406 \text{\AA}$). Magnetization measurements were performed using a Quantum Design Magnetic Property Measurement System (MPMS 3) using a vibrating sample magnetometer. Electrical and thermal transport measurements were performed using a Quantum Design Physical Property Measurement System (PPMS).

The powder XRD pattern confirms that ScVTiHfNb crystallizes in a cubic crystal structure (bcc) with the space group Im3$\bar{m}$ (229).  Le-Bail refinement was performed using FullProf Suite software \cite{le1988ab}. The refined powder XRD is shown in \figref{Fig1}(a), which yields the lattice parameter $a=3.3267(7)\ \text{\AA}$. Energy-dispersive X-ray diffraction (EDX) at various locations in the sample reveals the average phase composition as Sc$_{0.18}$V$_{0.20}$Ti$_{0.20}$Hf$_{0.19}$Nb$_{0.23}$ which is close to the nominal composition within the experimental error limit (<5\%). Furthermore, the EDX elemental mapping shows the uniform distribution of the elements Sc, V, Ti, Hf, and Nb, indicating a homogeneous phase formation as shown in \figref{Fig1}(b). \figref{Fig1}(c) illustrates the crystal structure of ScVTiHfNb, where the Wyckoff position can be occupied by any element, resulting in site mixing.


\begin{figure*}
\includegraphics[width=1.99\columnwidth, origin=b]{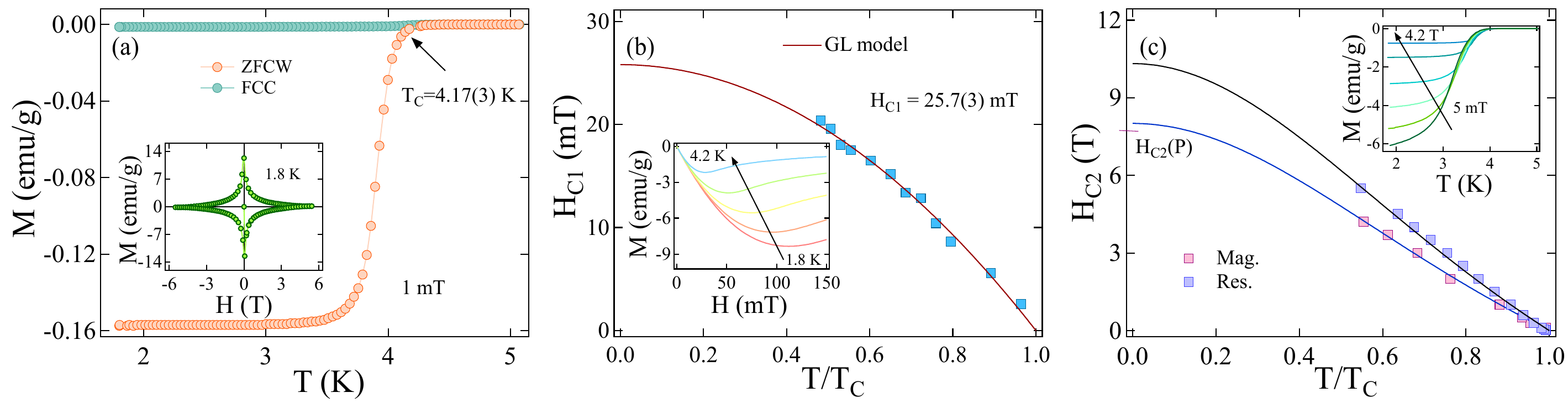}
\caption{\label{Fig2}(a) Temperature variation of magnetization in the ZFCW and FCC modes shows the appearance of superconductivity at 4.17(3) K. The inset shows the field-dependent magnetization loop at 1.8 K, revealing the type-II nature of superconductivity. (b) Temperature dependence of the lower critical field H$_{C1}(T)$ fitted with \equref{eqn2:Hc1} (solid red line). The inset shows isothermal magnetization vs field measurements at various temperatures. (c) Temperature variation of the upper critical field H$_{C2}(T)$ calculated from magnetization vs. temperature curves (pink squares) and resistivity vs. temperature curves (purple triangles) fitted with \equref{eqn3:Hc2}. The inset shows the magnetization vs. temperature curves at various applied magnetic fields.}
\end{figure*}

Electrical resistivity measurements were performed from 1.9 to 300 K in the zero field for ScVTiHfNb, as shown in \figref{Fig1}(d). The electrical resistivity abruptly drops to zero as the temperature decreases, exhibiting superconductivity with a transition temperature $T_C^{mid} = 4.47(1)$ K. \figref{Fig1}(e) shows the zero-field electrical resistivity in the vicinity of the superconducting transition. After $T_C$, the resistivity gradually increases with increasing temperature, indicating a poor metallic nature. The residual resistivity ratio (RRR) is estimated to be $\rho_{300 K}/\rho_{8 K} = 1.06(2)$, implying strong electron scattering and a highly disordered nature, which can also be seen in other high-entropy alloy superconductors \cite{motla2021probing, motla2022boron, motla2022superconducting, motla2023superconducting}. The normal-state electrical resistivity is well-fitted with the following expression:
\begin{equation}
    \rho(T)=\rho_0+C\left(\frac{T}{\theta_\mathrm{D}}\right)^n \int^{\theta_\mathrm{D}/T}_0 \frac{x^n}{\left(e^x-1\right)\left(1-e^{-x}\right)}\,dx.
    \label{eqn1:BG}
\end{equation}
Here, the first term $\rho_0$ is the residual resistivity, and the second is the Bloch-Gr\"{u}neisen expression \cite{Grimvall1981electron}. $C$ is a material-dependent property, whereas $n$ depends on the nature of the interaction, and $\theta_\mathrm{D}$ is the Debye temperature \cite{bid2006temperature}. The best fit is obtained for $n=3$, which yields $\rho_0 = 152.77(5)\ \mu\Omega$-cm and $\theta_\mathrm{D}=224(12)$ K. 
The Kadowaki-Woods ratio ($K_w = A/\gamma^2_n$) quantifies the strength of the electron-electron correlation \cite{kadowaki1986universal}, where the coefficient $A$ represents electron-electron scattering at low temperatures that contribute to electronic resistivity, while $\gamma_n$ represents the Sommerfeld coefficient (obtained from the specific heat measurement). The coefficient $A$ is obtained by fitting the data of the normal state resistivity with the power law ($\rho=\rho_0+AT^2$) at low temperatures (shown in \figref{Fig1}(d)). Using $A=3.6(4)\times10^{-4}\ \mu\Omega$-cmK$^{-2}$ and $\gamma_n=7.41(7)$ mJ-mol$^{-1}$K$^{-2}$, the resultant value of $K_w$ is $0.65(8)\times10^{-5}\ \mu\Omega$-cmK$^{2}$mJ$^{-2}$mol$^{2}$, which is less than $1\times10^{-5}\ \mu\Omega$-cmK$^{2}$mJ$^{-2}$mol$^{2}$ categorizing it as a weakly correlated system. In addition to the zero-field measurement, resistivity measurements at different applied magnetic fields, as shown in \figref{Fig1}(f), were also performed to evaluate the upper critical field, discussed in the next section. Furthermore, the Hall resistivity
($\rho_{xy}(H)$) is also measured to extract the carrier concentration, which is found to be $n = 1.8(1)\times10^{29}$ m$^{-3}$.


The temperature variation of DC magnetization in both zero-field-cooled-warming (ZFCW) and field-cooled-cooling (FCC) modes at a 1 mT applied magnetic field shows a transition into a diamagnetic state below 4.17(3)K, marked as $T_C$ as shown in \figref{Fig2}(a). The difference between the ZFCW and FCC modes below $T_C$ is due to the strong flux pinning. The inset of \figref{Fig2}(a) represents the magnetization loop at 1.8 K, revealing the type-II nature of superconductivity in ScVTiHfNb. Field-dependent magnetization measurements were also performed at different temperatures ranging from 1.8 to 4.2 K (shown in the inset of \figref{Fig2}(b)) to extract the lower critical field $H_{C1}(T)$. $H_{C1}$ for each temperature was taken as the point at which the M-H curve deviates from linearity. The behavior of $H_{C1}(T)$ is well described by the Ginzburg-Landau (GL) equation as:
\begin{equation}
    H_{C1}(T)=H_{C1}(0)\left[1-\left(\frac{T}{T_{C}}\right)^{2}\right].
    \label{eqn2:Hc1} 
\end{equation}
\figref{Fig2}(b) shows the temperature dependence of lower critical field $H_{C1}(T)$, which is fitted with \equref{eqn2:Hc1} and yields $H_{C1}(0)$ = 25.7(3) mT by extrapolating the fit to 0 K. Furthermore, magnetization vs. temperature measurements (inset of \figref{Fig2}(c)) and resistivity vs. temperature measurements (\figref{Fig1}(f)) were performed in various applied fields to calculate the upper critical field ($H_{C2}$). $H_{C2}$ was evaluated by the change in observed $T_C$ for every field, since $T_C$ decreases when the applied field increases. The temperature evolution of the upper critical field $H_{C2}(T)$ is fitted by the Ginzburg-Landau (GL) equation, as shown in \figref{Fig2}(c),
\begin{equation}
H_{C2}(T) = H_{C2}(0)\left[\frac{(1-t^{2})}{(1+t^2)}\right]
\label{eqn3:Hc2}
\end{equation}
where $t=T/T_C$ is the reduced temperature. The estimated values of $H_{C2}(0)$ by extrapolating the GL fit up to 0 K from magnetization and resistivity measurements are 7.9(2) T and 10.3(1) T, respectively. The GL coherence length $\xi_{\mathrm{GL}}(0)$ can be evaluated using the upper critical field as
$H_{C2}(0) = \frac{\phi_0}{2\pi \xi^2_{\mathrm{GL}}(0)}$,
where $\phi_0 = 2.07\times10^{-15}$ T-m$^2$ is the magnetic flux quantum \cite{Tinkham}. By substituting the values of $H_{C2}(0)$ and $\phi_0$, the GL coherence length is estimated to be $\xi_{\mathrm{GL}}(0)=64(2)$ \text{\AA}.

The value of $\xi_{\mathrm{GL}}(0)$ can further be used to calculate another characteristic length, penetration depth which is related to $H_{C1}(0)$ via the expression \cite{klimczuk2007physical}
\begin{equation}
    H_{C1}(0) = \frac{\phi_0}{4\pi \lambda^2_{\mathrm{GL}}(0)}\left(\ln{\frac{\lambda_{\mathrm{GL}}(0)}{\xi_{\mathrm{GL}}(0)}+0.12}\right).
    \label{eqn4:lambda_GL}
\end{equation}
The estimated value of $\lambda_{\mathrm{GL}}(0)$ is 1437(62) \text{\AA}. The GL parameter ($\kappa_{\mathrm{GL}}$) is calculated as $\kappa_{\mathrm{GL}}=\frac{\lambda_{\mathrm{GL}}(0)}{\xi_{\mathrm{GL}}(0)}= 22(2)>>1/\sqrt{2}$, further confirming type-II superconductivity in ScVTiHfNb.
The thermodynamic critical field is determined by the values of $\kappa_{\mathrm{GL}}$, $H_{C1}(0)$ and $H_{C2}(0)$ using the expression $H_{C1}(0)H_{C2}(0)=H^2_C(0)\ln{\kappa_{\mathrm{GL}}}$ yielding $H_C(0)=256(2)$ mT \cite{klimczuk2007physical}.

The superconductivity can be killed by applying an external magnetic field greater than the upper critical field. In type-II superconductors, it can be done via two mechanisms: the orbital limiting effect and the Pauli paramagnetic effect. The orbital limiting field can be described by the Werthamer-Helfand-Hohenberg (WHH) model in the weak coupling limit as \cite{werthamer1966, helfand1966}:
\begin{equation}
    H_{C2}^{\mathrm{orb}}(0) = -\alpha T_C \left. \frac{dH_{C2}(T)}{dT}\right|_{T=T_{C}}.
    \label{eqn5:Hc2_orb}
\end{equation}
Here $\alpha = 0.693$ for BCS dirty limit superconductor. Substituting the value of $\left. T_C \frac{dH_{C2}(T)}{dT}\right|_{T=T_{C}} = -8.2(8)$ T, the estimated value of $H_{C2}^{\mathrm{orb}}(0)$ is 5.7(5) T. The Pauli paramagnetic field in the BCS weak coupling limit can be expressed as $H_{C2}^{\mathrm{P}}(0) = 1.86 T_C$ \cite{chandrasekhar1962, clogston1962}. Taking $T_C=4.17(3)$ K, $H_{C2}^{\mathrm{P}}(0)$ is calculated to be 7.75(5) T. The Maki parameter expressed as $\alpha_\mathrm{M}=\sqrt{2} H_{C2}^{\mathrm{orb}}(0)/H_{C2}^{\mathrm{P}}(0)$ is used to find the relative strength of the orbital limiting effect to the Pauli paramagnetic effect, and it is found to be 1.05(9) for ScVTiHfNb \cite{maki1966}. According to the value of $\alpha_\mathrm{M}$, the pair breaking is mainly caused by the Pauli paramagnetic effect with a small contribution of the orbital limiting effect.

The Ginzburg number $G_i$, expressed as the ratio of the thermal energy ($k_\mathrm{B}T$) to the condensation energy, which is connected with coherence volume, is given as \cite{blatter1994vortices}:
\begin{equation}
    G_i = \frac{1}{2}\left(\frac{k_\mathrm{B}\mu_0\tau T_C}{4\pi\xi^3_{\mathrm{GL}}(0)H^2_C(0)}\right)^2.
\end{equation}
The anisotropic ratio $\tau$ is 1 for cubic ScVTiHfNb. By substituting the values $\xi_{\mathrm{GL}}(0)=64(2)$ \AA, $H_C(0)=256(2)$ mT and $T_C=4.17(3)$ K, we obtained $G_i=5.58(5)\times10^{-8}$ which is comparable to that of the low-temperature superconductors.

\begin{figure}
\includegraphics[width=0.90\columnwidth, origin=b]{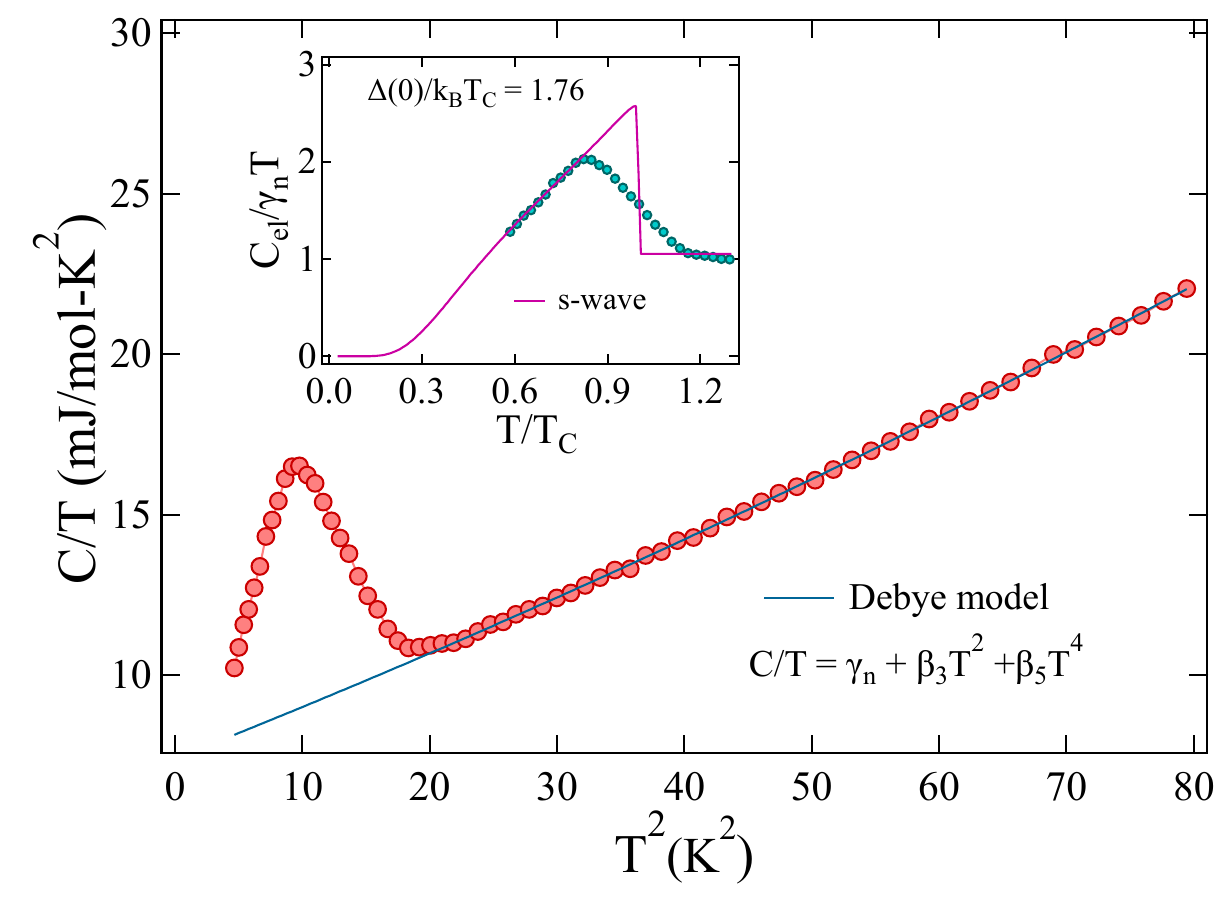}
\caption{\label{Fig3}The electronic specific heat with respect to temperature is well fitted to the BCS s-wave model (solid pink line). The inset shows C/T vs. T$^2$ at zero fields, which fits well with the Debye model (solid green line).}
\end{figure}
The specific heat is also performed to characterize the thermal properties of ScVTiHfNb in zero field, shown in \figref{Fig3}. A transition from normal to superconducting state is observed at 3.72(1) K, associated with a pronounced jump in specific heat. The slight variation of $T_C$ from the magnetization and resistivity measurements could be due to the broadness of the jump in the specific heat of the sample \cite{kasem2021anomalous}. In the normal state, $C(T)/T$ is best fitted with the Debye model $\frac{C(T)}{T} = \gamma_n +\beta_3T^2 + \beta_5T^4$, where $\gamma_n$ is the Sommerfeld coefficient, $\beta_3$ is the phononic contribution to the specific heat, and $\beta_5$ is the anharmonic contribution. By extrapolating the fit to 0 K, $\gamma_n$, $\beta_3$ and $\beta_5$ are estimated to be 7.41(7) mJ-mol$^{-1}$K$^{-2}$, 0.157(3) mJ-mol$^{-1}$K$^{-4}$ and 0.34(3) $\mu$J-mol$^{-1}$K$^{-6}$, respectively. The Debye temperature is related to $\beta_3$ through the expression
$ \theta_\mathrm{D} = \left(\frac{12\pi^4RN}{5\beta_3}\right)^{\frac{1}{3}}$ \cite{Kittel},
where $R=8.31$ Jmol$^{-1}$K$^{-1}$ is the universal gas constant, and $N$ is the number of atoms per formula unit, which is 1 for ScVTiHfNb. The estimated value of $\theta_\mathrm{D}$ is 231(1) K after substituting the values of $R$, $N$, and $\beta_3$. For a non-interactive system, the density of states at the Fermi level [$D_C(E_\mathrm{F})$] has a direct impact on the Sommerfeld coefficient as $\gamma_n = \left(\frac{\pi^2k^2_\mathrm{B}}{3}\right)D_C(E_\mathrm{F})$.
Here $k_\mathrm{B}=1.38\times10^{-23}$ JK$^{-1}$ is Boltzmann's constant, and substituting the value of $\gamma_n$, $D_C(E_\mathrm{F})$ is estimated to be 3.14(3) states/eV-f.u. The electron-phonon coupling constant $\lambda_{e-ph}$, which is related to $\theta_\mathrm{D}$ and $T_C$, was introduced by McMillan as a way to assess the strength of the coupling between the electron and the phonon, expressed as:
\begin{equation}
\lambda_{e-ph} = \frac{1.04+\mu^{*}\ln{\left(\theta_{\textrm{D}}/1.45T_{C}\right)}}{\left(1-0.62\mu^{*}\right)\ln{\left(\theta_{\textrm{D}}/1.45T_{C}\right)}-1.04},
\label{eqn6:e-ph}
\end{equation}
where $\mu^{*}$ is the screened Coulomb potential, assumed to be 0.13 for intermetallic compounds \cite{mcmillan1968transition}. Inserting the values of $\theta_\mathrm{D}$ and $T_{C}$, $\lambda_{e-ph}$ is evaluated to be 0.633(4), which indicates that ScVTiHfNb is a weakly coupled superconductor. The electronic specific heat at low temperature is calculated by deducting the lattice-specific heat from the total specific heat, which yields the specific heat jump $\Delta C_{el}/\gamma_nT_c=1.33(1)$ close to the BCS value of 1.43. The below expression of the s-wave isotropic fully gaped BCS model provided the best fit (shown in the inset of \figref{Fig3}) for the low-temperature electronic specific heat $C_{el}(T)$ data for normalized entropy S,
\begin{equation}
\frac{S}{\gamma_{n}T_{C}} = -\frac{6}{\pi^2}\left(\frac{\Delta(0)}{k_\mathrm{B}T_{C}}\right)\int_{0}^{\infty}\left\{f\ln{f}+(1-f)\ln{1-f}\right\}dy,
\label{eqn10:s-wave}
\end{equation}
where $f(\xi)=\left\{\exp\left(E(\xi)/k_\mathrm{B}T)\right)+1\right\}^{-1}$ is the Fermi function, $E(\xi)=\sqrt{\xi^{2}+\Delta^{2}(t)}$ is the energy of the normal electrons, $y=\xi/\Delta(0)$, $t=T/T_{C}$ and $\Delta(t)=\tanh{\left\{1.82(1.018((1/t)-1)^{0.51}\right\}}$ is the approximated BCS gap value. The electronic specific heat in the superconducting region is correlated with the normalized entropy by the relation $\frac{C_{el}}{\gamma_{n}T_{C}} = t\frac{d(S/\gamma_{n}T_{C})}{dt}$.
The value of the superconducting gap $\Delta(0)/k_\mathrm{B}T_C=1.76(6)$ is obtained by fitting the electronic specific heat data with \equref{eqn10:s-wave}, which is close to the BCS value 1.76, suggesting that ScVTiHfNb exhibits weakly coupled BCS superconductivity \cite{padamsee1973quasiparticle}.

\begin{figure}
\includegraphics[width=0.90\columnwidth, origin=b]{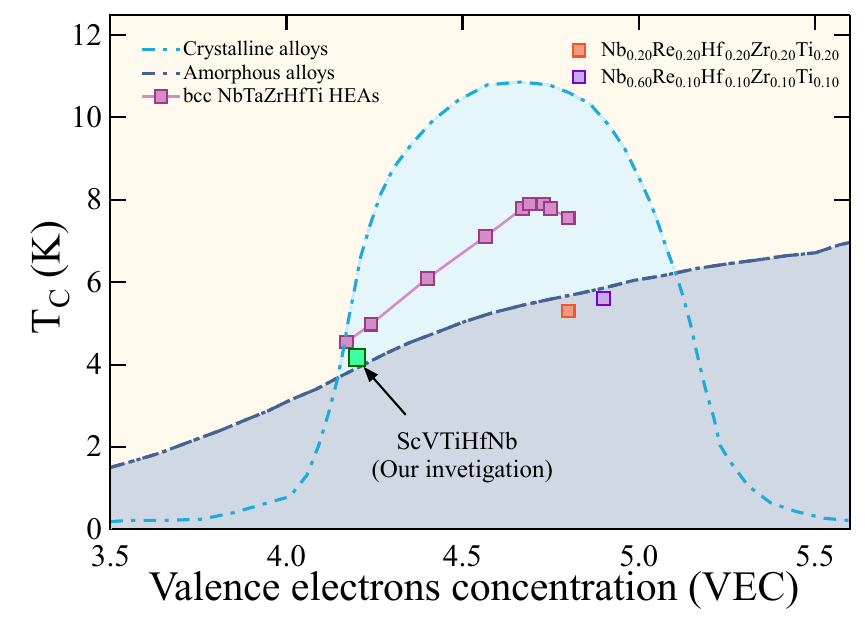}
\caption{\label{Fig4}The transition temperature dependence on the valence electron concentration for ScVTiHfNb with reference to the crystalline 4d metals, amorphous 4d metals, and other bcc HEAs \cite{matthias1955empirical, collver1973superconductivity, von2016effect, marik2018superconductivity, motla2022superconducting}.}
\end{figure}

The relationship between $T_C$ and the valence electron concentration (VEC) is illustrated in \figref{Fig4}. Comparative data from various sources, including studies on crystalline metals, amorphous metals, and other bcc HEAs, are incorporated for analysis \cite{matthias1955empirical, collver1973superconductivity, von2016effect, marik2018superconductivity, motla2022superconducting}. According to Rohr \textit{ et al.}, the VEC dependence of $T_C$ falls between crystalline and amorphous alloys, aligning with other similar VEC-valued HEA superconductors.

To validate the experimental findings of ScVTiHfNb, electronic properties are also extracted using experimentally evaluated parameters such as carrier concentration ($n$), Sommerfeld coefficient ($\gamma_n$), and residual resistivity ($\rho_0$). The Sommerfeld coefficient is directly related to the effective mass ($m^*$) and carrier concentration ($n$) of the quasiparticles via the relation $\gamma_{n} = \left(\frac{\pi}{3}\right)^{2/3}\frac{k_\mathrm{B}^{2}m^{*}n^{1/3}}{\hbar^{2}}$,
where $\hbar=1.05\times10^{-34}$ Js is the reduced Planck constant. Inserting the values of $\gamma_n=7.41(7)\ \textrm{mJ-mol}^{-1}\textrm{K}^{-2}$ and $n=1.8(1)\times10^{29}\ \textrm{m}^{-3}$ (from the Hall measurement), the effective mass $m^*$ is evaluated as $7.3(2)\ m_e$.
 Fermi velocity $v_\textrm{F}$ depends on $m^*$ and $n$ by relation $n=\frac{1}{3\pi^2}\left(\frac{m^*v_\textrm{F}}{\hbar}\right)^3$.
After substituting the values of $m^*$ and $n$, $v_\textrm{F}=2.7(1)\times10^5 \ \textrm{ms}^{-1}$ is obtained. The mean free path ($l$) is directly related to $\rho_0$, $m^*$ and $v_\textrm{F}$ through the expression $l = \frac{3\pi^2\hbar^3}{e^2\rho_0m^{*2}v^2_{\textrm{F}}}$. The estimated value of the mean free path is $l=2.6(3) \ \text{\AA}$, which is relatively low and agrees with other high-entropy alloy superconductors \cite{motla2021probing, motla2022superconducting, motla2023superconducting}. The high amount of disorder by five distinct elements in the crystal structure of ScVTiHfNb causes a low value of the mean free path. In the BCS framework, the coherence length ($\xi_0$) can be defined as $\xi_0=\frac{0.18\hbar v_{\textrm{F}}}{k_\textrm{B}T_C}$.
Substituting the values of $v_\textrm{F}$ and $T_C$, $\xi_0$ is found to be 909(7) \AA. The ratio of the BCS coherence length and the mean free path ($\xi_0/l$) classifies superconductors into clean or dirty limit superconductors. For ScVTiHfNb, $\xi_o>>l$, classifying it as a dirty limit superconductor.

\begin{figure}
\includegraphics[width=0.87\columnwidth, origin=b]{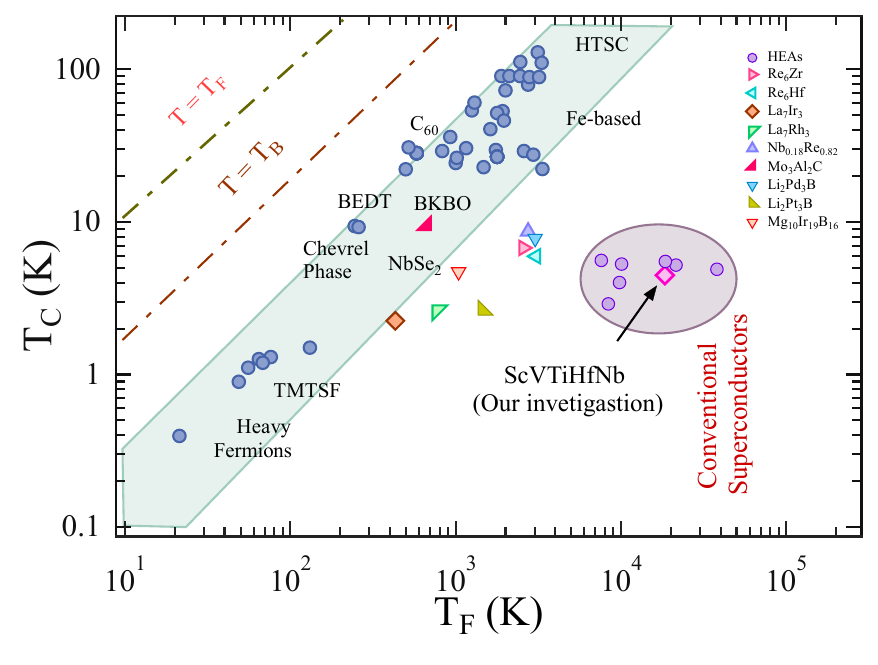}
\caption{\label{Fig5}Uemura plot between transition temperature ($T_C$) and Fermi temperature ($T_{\textrm{F}}$). The unconventional region is defined as $0.01\leq\frac{T_C}{T_\text{F}}\leq0.1$.}
\end{figure}

Uemura \textit{et al.} classified superconductors as conventional or unconventional, depending on the ratio of $T_C$/$T_\textrm{F}$ \cite{uemura1988systematic, uemura1989universal, uemura1991basic}. Unconventional superconductors exhibit ratios between 0.01 and 0.1, while conventional superconductors have $T_C/T_\textrm{F} \geq 0.1$. The Fermi temperature of a 3D system can be expressed as $k_\textrm{B}T_\textrm{F} = \frac{\hbar^2}{2m^*}\left(3\pi^2n\right)^{\frac{2}{3}}$ \cite{hillier1997classification}.
This equation provides $T_\textrm{F}=1.84(4)\times10^4$ K after substituting the values of $m^*$ and $n$. The ratio $T_C/T_\textrm{F}$ is 0.00022(6), which isolates it from the unconventional band represented by a green region, as shown in \figref{Fig5}. All parameters of the normal and superconducting state of ScVTiHfNb are summarized in \tableref{Tab2}.
\begin{table}
\caption{Parameters in the superconducting and normal state of ScVTiHfNb}
\label{Tab2}
\begin{center}
\setlength{\tabcolsep}{7 pt}
\begin{tabular}{l l l} 
\hline\hline
Parameters & Unit & Value  \\
\hline
VEC & & 4.2\\
$T_{C}$& K& 4.17(3)\\             
$H_{C1}(0)$& mT& 25.7(3)\\                       
$H_{C2}^{mag, res}(0)$& T& 7.9(2), 10.3(1)\\
$H_{C2}^{\textrm{P}}(0)$& T&7.75(5)\\
$H_{C2}^{orb}(0)$& T& 5.7(5) \\
$\xi_{\textrm{GL}}(0)$& \text{\AA}& 64(2)\\
$\lambda_{\textrm{GL}}(0)$& \text{\AA}& 1437(62)\\
$k_{\textrm{GL}}$& &22(2)\\
$\gamma_{n}$&  mJ-mol$^{-1}$K$^{-2}$& 7.41(7) \\   
$\theta_\textrm{D}$& K& 231(1)\\
$\lambda_{e-ph}$& &0.633(4)\\
$\Delta C_{el}/\gamma_nT_C$ & &1.33(1)\\
$\Delta(0)/k_\textrm{B}T_C$& &1.76(6)\\
$\xi_{0}/l_{e}$& &  349(43)\\
$v_{\textrm{F}}$& 10$^{5}$ ms$^{-1}$& 2.7(1)\\
$n$& 10$^{29}$m$^{-3}$& 1.8(1)\\
$T_{\textrm{F}}$&10$^{4}$ K& 1.84(4)\\
$m^{*}$/$m_{e}$&  & 7.3(2)\\
\hline\hline
\end{tabular}
\end{center}
\end{table}
  
In summary, we have synthesized a high-entropy equiatomic alloy, ScVTiHfNb, which exhibits a body-centered cubic (bcc) crystal structure based on the Hume-Rothery rule (VEC = 4.2). This position it within the region of bcc superconducting high-entropy alloys (HEAs). Our examination, which includes magnetization, electrical resistivity, and specific heat measurements, delves into the properties of both normal and superconducting states. Our comprehensive findings reveal bulk superconductivity with a transition temperature of approximately 4.17(3) K. Heat capacity measurements indicate weakly coupled, fully gapped superconductivity. The upper critical field, extrapolated from magnetization and resistivity data, surpasses the Pauli paramagnetic field, suggesting potential unconventional behavior. The equiatomic composition of HEAs, devoid of high atomic concentrations of specific elements, holds promise for the development of superconductors with elevated upper-critical fields.
Additionally, equiatomic superconducting high-entropy alloys, known for their maximal disorder, offer a distinct opportunity to explore the superconducting pairing mechanism more clearly. Their metallic nature, moderate transition temperature, high critical field, and possibility of conversion to thin film form make them suitable for device applications. To deepen our understanding, comprehensive studies, including microscopic techniques such as muon spin rotation/relaxation measurement and theoretical investigations of the electronic structure, can elucidate the pairing mechanism in these materials.

\section{Acknowledgement}
R.P.S. acknowledges the
SERB Government of India for the Core Research Grant CRG/2023/000817.

\pagestyle{plain}
\addcontentsline{toc}{chapter}{Bibliography}

\end{document}